\documentclass[]{interact}

\usepackage{epstopdf}
\usepackage[caption=false]{subfig}

\usepackage{graphicx}
\usepackage{mathtools}
\usepackage{commath}
\usepackage{indentfirst}
\usepackage{graphicx}
\usepackage[dotinlabels]{titletoc}
\usepackage{xcolor}
\usepackage[pagewise]{lineno}

\usepackage[numbers,sort&compress]{natbib}
\bibpunct[, ]{[}{]}{,}{n}{,}{,}
\makeatletter
\def\NAT@def@citea{\def\@citea{\NAT@separator}}
\makeatother

\theoremstyle{plain}

\theoremstyle{definition}

\theoremstyle{remark}

\begin{document}


\title{A Robust Adaptive Modified Maximum Likelihood Estimator for the Linear Regression Model}

\author{
\name{S.~Acitas\textsuperscript{a}\thanks{CONTACT S. Acitas. Email: sacitas@eskisehir.edu.tr};  Peter Filzmoser \textsuperscript{b}  Birdal Senoglu \textsuperscript{c}}
\affil{\textsuperscript{a}Eskisehir Technical University, Department of Statistics, Eskisehir, Turkey; \textsuperscript{b}Vienna University of Technology, Vienna, Austria; \textsuperscript{c}Ankara University, Department of Statistics, Ankara, Turkey}
}

\maketitle

\begin{abstract}
In linear regression, the least squares (LS) estimator has certain optimality properties if the errors are normally distributed. This assumption is often violated in practice, partly caused by data outliers. Robust estimators can cope with this situation and thus they are widely used in practice. One example of robust estimators for regression are adaptive modified maximum likelihood (AMML) estimators \cite{Donmez2010}. However, they are not robust to $x$ outliers, so-called leverage points. In this study, we propose a new regression estimator by employing an appropriate weighting scheme in the AMML estimation method. The resulting estimator is called  robust AMML (RAMML) since it is not only robust to $y$ outliers but also to $x$ outliers.  A simulation study is carried out to compare the performance of the RAMML estimator with some existing robust estimators such as MM, least trimmed squares (LTS) and S. The results show that the RAMML estimator is preferable in most settings according to the mean squared error (MSE) criterion.  Two data sets taken from the literature are also analyzed to show the implementation of the RAMML estimation methodology.  
\end{abstract}

\begin{keywords}
Adaptive modified likelihood; efficiency;  leverage point;  regression;  robustness
\end{keywords}

\section{Introduction}

Regression analysis is a widely used statistical tool to analyze the relationships  between a dependent variable and predictors.  The least squares (LS) method is usually utilized to estimate the unknown model parameters. The LS estimator is popular because of its
simple computation and good statistical properties, i.e.~it is the best linear unbiased estimator under the normality assumption. 

The LS estimator loses its efficiency when the normality assumption is not satisfied. Therefore, some alternative methods are employed to accommodate the problems arising in case of non-normality which is caused by outliers.  Indeed, there are two types of outliers in the context of regression analysis: (i) vertical outliers, and (ii) leverage points. These outliers refer to outlyingness in the space of the response and the predictors, respectively.
The LS estimator is sensitive  to both types of outliers, and bad leverage points 
can even distort the estimator and lead to a non-sense model \cite{RousseeuwLeroy1987}. For this reason, many robust estimators which reduce the effects of outliers have been proposed in the literature. For example, the M estimator \cite{Huber1973}, the S estimator \cite{RousseeuwYohai1984}, the least trimmed squares (LTS) and the least median of squares (LMS) of  Rousseeuw \cite{Rousseeuw1984}, and the MM estimator \cite{Yohai1987} are well-known and widely used in the context of robust regression. We refer to Maronna et al. \cite{Maronnaetal2006} or Gschwandtner and Filzmoser \cite{GschwandtnerFilzmoser2012} in which descriptions of these estimators and related literature information are briefly given. See also Yu and Yao \cite{YuYao2017} for an extensive review of robust linear regression estimators. 

In the literature, the modified maximum likelihood (MML) estimator is also used to reduce the effect of outlying observations. As its name indicates, the MML estimator is obtained applying a particular modification of the maximum likelihood (ML) method \cite{Tiku1967,Tiku1968}.  This modification allows to obtain explicit forms of the estimator under the assumption of a non-normal error distribution. A long-tailed symmetric distribution is frequently used in this context since it is useful for modelling outliers occurring in the direction of long tails \cite{TikuSuresh1992}. For example, Tiku et al.~\cite{Tikuetal2001} obtain the MML estimator of the unknown parameters of the simple linear regression model when the distribution of the error terms is long-tailed symmetric. The MML estimator is also obtained for the multiple linear regression model \cite{IslamTiku2005} with long-tailed symmetric error distributions. The MML estimator has good properties, and it is easy to compute.  Furthermore, it is not only robust to outliers but also asymptotically equivalent to the ML estimator. However, it has two drawbacks: (i) the  shape parameter of the long-tailed symmetric distribution is assumed to be known and (ii) it is only robust to vertical outliers but not to leverage points. 

The first drawback is not a new issue in the related literature. For example, Lucas \cite{Lucas1997} consider the Student's $t$ distribution as an alternative to normal distribution for estimating the location parameter under the assumption of known degrees of freedom $\nu$. This is because of the fact that the ML estimator of the other parameters is no longer robust when it is estimated jointly with $\nu$. Therefore, it is treated as a robustness tuning constant-- see also Arslan and Genc \cite{ArslanGenc2003,ArslanGenc2009}, Acitas et al. \cite{Acitasetal2013a,Acitasetal2013b} in which similar discussions are available. Tiku and Surucu \cite{TikuSurucu2012} also consider this issue and propose a new version of the MML estimator. This version is then called adaptive MML (AMML) by Donmez \cite{Donmez2010}. In AMML, the main idea of the MML methodology is preserved but the assumption of a known shape parameter on the estimation process is weakened.  

To the best of our knowledge, the second drawback is still an open problem. In other words, there is no previous study on the MML and AMML estimators dealing with leverage points as far as we know. Therefore, the motivation of this study is to suggest a new regression estimator which is robust to both types of outliers based on the AMML estimators. This estimator is called robust AMML (RAMML) and obtained by introducing new weights assigned to the predictors in the AMML estimation process \cite{Acitasetal2019}. The RAMML is not only robust to both types of outliers but also efficient under the normality assumption. A further advantage of the  RAMML estimator is that it is easy to compute since the explicit form of the estimator is available.

The rest of the paper is organized as follows. Section  \ref{sec:amml} reviews the AMML methodoloy. Section \ref{sec:ramml} is reserved to the RAMML estimator. 
The results of a Monte-Carlo simulation study are given in Section \ref{sec:simulation}. The paper is finalized with some concluding remarks.

\section{Review of the AMML estimator} \label{sec:amml}


We consider a multiple linear regression model with $n$ observations and $m$ predictors,
\begin{equation}\label{eq:linearmodel}
y_i =\beta_0+ \mathbf{x}_i'\boldsymbol{\beta}+\varepsilon_i, \quad i=1,2,\ldots,n
\end{equation}
where $y_i$ is the dependent variable, $\beta_0$ is the intercept,  $\boldsymbol{\beta}=(\begin{matrix}
\beta_1 & \beta_2 & \cdots & \beta_m \\
\end{matrix})' $ is the vector of unknown regression coefficients, $\mathbf{x}_i$ is the vector of predictors defined by  $\mathbf{x}_i=\left(\begin{matrix}
x_{i1} & x_{i2} & \cdots & x_{im} \\
\end{matrix}\right)'$  and $\varepsilon_i$  is the random error term.

Assume that the distribution of the error terms in model \eqref{eq:linearmodel} is long-tailed symmetric with the probability density function (pdf) given by
\begin{equation}
f(e;p,\sigma)=\frac{1}{\sqrt{q}B(0.5,p-0.5)\sigma}\left(1+\frac{e^2}{q\sigma^2}\right)^{-p},\quad q=2p-3, \quad -\infty<e<\infty
\end{equation}
where $p$ is the shape parameter which is assumed to be $p\geq 2$, $\sigma$ is the scale parameter and $B(\cdot,\cdot)$ is the beta function. The kurtosis value of the long-tailed distribution is greater than 3.  It tends to normal distribution as $p$ tends to $\infty$. Therefore, it has been considered as a good alternative to the normal distribution \cite{TikuSuresh1992}.

The ML estimators of the model parameters are the solutions of the following likelihood equations: 
\begin{eqnarray}
\label{mulmmlol1}\frac{\partial \log L}{\partial \beta_0}&=& \frac{2p}{q\sigma}\sum_{i=1}^ng(z_i)=0, \\
\label{mulmmlol2}\frac{\partial \log L}{\partial \beta_j}&=& \frac{2p}{q\sigma}\sum_{i=1}^ng(z_i)x_{ij}=0, \quad
j=1,2,\ldots,m\\
\label{mulmmlol3}\frac{\partial \log L}{\partial \sigma}&=& -\frac{n}{\sigma}+\frac{2p}{q\sigma}\sum_{i=1}^n
g(z_i) z_{i}=0 ,
\end{eqnarray}
where
\[\  g(z_i)=\frac{z_{i}}{1+\displaystyle\frac{z_{i}^2}{q}}, \quad   z_{i}=\displaystyle \frac{y_i-\beta_0-\mathbf{x}'_i\boldsymbol{\beta}}{\sigma}, \quad i=1,2,\ldots,n. \]
Since $g(\cdot)$  is a nonlinear function of the unknown model parameters, the explicit solutions of equations \eqref{mulmmlol1}-\eqref{mulmmlol3} cannot be obtained,
and numerical methods need to be used instead.
However, using numerical methods has some drawbacks such as wrong convergency, multiple roots and non-convergence of iterations, see e.g. Puthenpura and Sinha \cite{PuthenpuraSinha1986} ,  Vaughan \cite{Vaughan1992}. The modified maximum likelihood (MML) methodology proposed in Tiku \cite{Tiku1967,Tiku1968} overcomes these problems and allows to obtain the explicit forms of the estimators.  The idea underlying the MML methodology is linearization  of the intractable term (i.e., the function $g(\cdot)$) in the likelihood equations. This is done using the first two terms of a Taylor series expansion. 
For more details we refer to Islam and Tiku \cite{IslamTiku2005} in which MML estimators for the multiple linear regression model are obtained under a long-tailed symmetric error distribution assumption. 

The MML estimators are explicitly formulated, thus no computational efforts are required. They have also good statistical proporties, i.e.~they are asymptotically equivalent to the ML estimators and therefore they are minimum variance bound estimators \cite{Bhattacharyya1985}. However, the MML methodology works under the assumption of a known shape parameter $p$.  In some studies, $p$ is identified using the profile likelihood method, see for example Acitas et al. \cite{Acitasetal2013a,Acitasetal2013b}.  Tiku and Surucu \cite{TikuSurucu2012}  suggested a new version of the MML method in which  the known shape parameter  assumption is weakened. They estimate the location and the scale parameter using this new version and obtain estimators which are as good as or better than M-estimators. Donmez \cite{Donmez2010} generalized this idea to the simple and multiple linear regression model and calls the method AMML methodology. 
In the following, the steps of the AMML are explained based on Donmez \cite{Donmez2010}:

\begin{description}
	\item[Step 1.]  The function $g(\cdot)$ is linearized around 
	\begin{equation}
	t_i=\dfrac{y_i-\beta_0-\mathbf{x}_i' \boldsymbol{\beta}}{\sigma}, \quad i=1,2,\ldots,n
	\end{equation}
	using the first two terms of a Taylor series expansion,
	\begin{equation} \label{gfnk}
	g(z_{i})\cong \alpha_i+\delta_iz_{i}, \quad i=1,2,\ldots,n
	\end{equation}
	where
	\begin{equation}\label{at}
	\alpha_i=\displaystyle \frac{(1/q))t_{i}}{(1+(1/q)t_{i}^2)^2}, \quad \delta_i=\displaystyle \frac{1}{(1+(1/q)t_{i}^2)^2},\quad
	i=1,2,\ldots,n.
	\end{equation}
	\item[Step 2.] The linearized version of the function $g(z_{(i)})$ is incorporated  into the likelihood equation and the so-called modified likelihood equations are obtained as follows:
	\begin{eqnarray}
	\label{mulmmlol1x}\frac{\partial \log L^*}{\partial \beta_0}&=& \frac{2p}{q\sigma}\sum_{i=1}^n \left(\alpha_i+\delta_i z_{i}\right) =0 \\
	\label{mulmmlol2x}\frac{\partial \log L^*}{\partial \beta_j}&=& \frac{2p}{q\sigma}\sum_{i=1}^n \left(\alpha_i+\delta_i z_{i}\right) x_{ij}=0, \quad
	j=1,2,\ldots,m\\
	\label{mulmmlol3x}\frac{\partial \log L^*}{\partial \sigma}&=& -\frac{n}{\sigma}+\frac{2p}{q\sigma}\sum_{i=1}^n
	\left(\alpha_i+\delta_i z_{i}\right)z_{i}=0 ,
	\end{eqnarray}
	where $\log L^*$stands for the modified log-likelihood function. 
	
	\item[Step 3.] The solutions of the Equations \eqref{mulmmlol1x}-\eqref{mulmmlol3x} are the AMML estimators. They are formulated as follows:
	\begin{eqnarray}
	\label{beta0mmllts} \hat{\beta}_{0} &=& \bar{y}_{[\cdot]}-\overline{\mathbf{x} }'_{[\cdot]}\boldsymbol{\hat\beta}\\
	\label{betammllts} \boldsymbol{\hat\beta}&=& \mathbf{K}+\mathbf{L}\hat\sigma,\\
	\label{sigmammllts}\hat\sigma&=& \frac{B+\sqrt{B^2+4nC}}{2\sqrt{n(n-m-1)}}
	\end{eqnarray}
	where
	\begin{eqnarray*}
		\mathbf{K}&=&(\mathbf{X}' \boldsymbol{\delta}\mathbf{X})^{-1}(\mathbf{X}' \boldsymbol{\delta}\mathbf{y}),\quad 	\mathbf{L}=(\mathbf{X}' \boldsymbol{\delta}\mathbf{X})^{-1}(\mathbf{X}' \boldsymbol{\alpha}\mathbf{1}) \\
		B&=&\frac{2p}{q}(\mathbf{y}-\mathbf{X}\mathbf{K})'\boldsymbol{\alpha} \mathbf{1} , \quad C=\frac{2p}{q}(\mathbf{y}-\mathbf{X}\mathbf{K})'\boldsymbol{\delta} (\mathbf{y}-\mathbf{X}\mathbf{K})
	\end{eqnarray*}
	\begin{eqnarray*}
		\bar{y}_{[\cdot]}= \frac{\displaystyle
			\sum_{i=1}^n \delta_i y_{i}}{\displaystyle \sum_{i=1}^n \delta_i}, 
		\quad \bar{\mathbf{x}}_{[\cdot] j}= \frac{\displaystyle\sum_{i=1}^n \delta_i x_{ij}}{\displaystyle\sum_{i=1}^n \delta_i},   \quad 
		\overline{\mathbf{x} }'_{[\cdot]}&=&[\bar{\mathbf{x}}_{[\cdot]1} \quad  \bar{\mathbf{x}}_{[\cdot]2} \quad  \cdots \quad  \bar{\mathbf{x}}_{[\cdot] m} ],
	\end{eqnarray*}
	\begin{eqnarray*}
		\boldsymbol{\delta}= diag(\delta_i) \quad \textrm{ and } \quad  \boldsymbol{\alpha}= diag(\alpha_i). 
	\end{eqnarray*}
	
	Here, $\mathbf{X}$ is the $n\times m$ matrix of predictors, and $\mathbf{1}$ stands for the vector of ones with $n$ entries.   
\end{description}

It is clear that AMML estimators are also explicitly formulated. They preserve the properties of the MML estimators \cite{Donmez2010}.  It can easily be shown that the AMML estimators are regression, scale and affine equvariant. The proof is not given here for the sake brevity, but it can be provided upon request. 

As it is clear, the values $t_i$ should be computed at the beginning of the methodology using some initial estimates. The computational details are explained in the following subsection. 

\subsection{Computation of the AMML estimators}

The computation of the AMML estimators is obtained in two iterations. An initial estimate of $t_i$ is required in the first iteration. The initial estimate of $t_i$, proposed by \cite{Donmez2010}, is obtained based on a repara\-metrization of the model \eqref{eq:linearmodel}. The reparametrized model is constructed assuming that all $\beta_j$ coefficients are equal 
in model \eqref{eq:linearmodel}. This is because of the fact that there is no reason to believe that one predictor is more important than others at the beginning. The reparametrized model is formulated by
\begin{equation}\label{eq:totalmodel}
y_i =\beta_0 +\theta v_i +\varepsilon_i, \quad v_i=\sum_{j=1}^{m} x_{ij}.  \quad i=1,2,\ldots,n.
\end{equation}
Then, the initial estimate of $t_{i}$ is given as 
\begin{equation}\label{eq:adaptivemmlt1}
\tilde{t}_{i}=\frac{y_{i}-T_0-T_1v_{i}}{S_0}
\end{equation}
where the initial estimators of $\beta_0$, $\theta$ and $\sigma$ are given by
\begin{equation}\label{eq:adaptiveinitial1}
T_0=median\{y_i-T_1v_i\}, \quad  T_1=median\left\{\frac{y_{\ell+1}-y_{\ell}}{v_{\ell+1}-v_{\ell}} \right\}   
\end{equation}
and
\begin{equation}\label{eq:adaptiveinitial2}
S_0=1.483 \cdot median\{  |y_{i}-T_0-T_1v_{i}|  \},    i=1,2,\ldots,n ,
\end{equation}
respectively.  After getting the initial estimate of $t_i$,  initial AMML estimators denoted by $\tilde{\beta}_{0}$, $\tilde{\beta}_{j}$ and $\tilde{\sigma}$ are obtained following steps 1-3 given in Section \ref{sec:amml}.

In the second iteration, $t_i$  is updated using the initial AMML estimators ($\tilde{\beta}_{0}$, $\tilde{\beta}_{j}$ and $\tilde{\sigma}$) as follows:
\begin{equation}
t_{i}=\frac{y_{i}-\tilde{\beta}_{0}-\displaystyle\sum_{j=1}^n x_{ij}\tilde{\beta}_{j}}{\tilde{\sigma}}
\end{equation}
and steps 1-3 are utilized one more time. The final AMML estimators are obtained at the end of  step 3 using Equations \eqref{beta0mmllts}-\eqref{sigmammllts}. 

It is clear that the shape parameter of the long-tailed symmetric distribution $p$ appears in the formulas. It is considered as robustness tuning constant and taken to be 16.5 so that the resulting AMML estimators are efficient when the distribution of the error terms is normal \cite{Donmez2010,TikuSurucu2012}.  

Using different simulation schemes, Donmez \cite{Donmez2010} shows that AMML estimators are robust to $y-$outliers. On the other hand, there are no avaliable results whether they are resistant to $x-$outliers. Indeed, the simulation schemes used in Donmez  \cite{Donmez2010} do not include any $x-$outliers. It should also be mentioned that Donmez  \cite{Donmez2010} point out to use standardized versions of predictors to reduce the effect of $x-$outliers. However, the standardization is done using mean and standard deviation which are not robust to outliers. Standardization is indeed required when working with the reparametrized model (see Equation \eqref{eq:totalmodel})  in which the sum of the predictors is used. If the predictors are measured under different scales and/or units, this sum is not remarkable. In our proposed methodology, we do not need any standardization for the predictors and the reparametrized model. Therefore, these AMML estimators are not considered in the rest of the paper. 

\section{Robust AMML estimators} \label{sec:ramml}
In this section, we propose RAMML estimators  which are robust to $x-$ outliers unlike 
the AMML. This is done by downweighting the $x-$outliers using extra weights for predictors.  The steps for the RAMML methodology are similar to those given for AMML,  i.e.~step 1 is 
the same while steps 2 and 3 have to be revised.  New versions of step 2 and 3 are shown by Step 2* and 3*, respectively, and explained below. 
\begin{description}
	\item[Step 2*.] Consider the modified likelihood Equations \eqref{mulmmlol1x}-\eqref{mulmmlol3x}. We introduce  new weights, denoted by $\delta^x_i$, for the predictors to accommodate the effects of $x$-outliers in these equations:
	\begin{eqnarray}
	\label{mulmmlol1xamml}\frac{\partial \log L^*}{\partial \beta_0}&=& \frac{2p}{q\sigma}\sum_{i=1}^n \delta^x_i \left(\alpha_i+\delta_i z_{i}\right) =0 \\
	\label{mulmmlol2xamml}\frac{\partial \log L^*}{\partial \beta_j}&=& \frac{2p}{q\sigma}\sum_{i=1}^n \delta^x_i \left(\alpha_i+\delta_i z_{i}\right) x_{ij}=0, \quad
	j=1,2,\ldots,m\\
	\label{mulmmlol3xamml}\frac{\partial \log L^*}{\partial \sigma}&=& -\frac{n}{\sigma}+\frac{2p}{q\sigma}\sum_{i=1}^n \delta^x_i
	\left(\alpha_i+\delta_i z_{i}\right)z_{i}=0 ,
	\end{eqnarray}
	where
	\begin{equation}\label{eq:atdeltax}
	\delta_i^x=\displaystyle \frac{1}{(1+(1/q) \tilde{x}_i^2)^4}\quad  \text{and} \quad 
	\tilde{x}_i=\frac{||\mathbf{x}_i-med_{L1}(\mathbf{X}) ||}{median_i||\mathbf{x}_i-med_{L1}(\mathbf{X}) ||}, \quad
	i=1,2,\ldots,n.
	\end{equation}
	Here, $med_{L1}(\cdot)$ denotes the $L_1-$median which is  also known as geometric median or spatial median. It is a highly robust estimator for the multivariate location, i.e.~its breakdown point is 0.5 \cite{Lopuhaaetal1991}.	We refer to  Croux et al.  \cite{Crouxetal2007} and Fritz et. al  \cite{Fritzetal2012} in which detailed information is available about $L_1-$median.
	\item[Step 3*.]  The RAMML estimators are obtained as solutions of 
	Equations~\eqref{mulmmlol1xamml}--\eqref{mulmmlol3xamml} . They are formulated as given in Equations~\eqref{beta0mmllts}--\eqref{sigmammllts}. However, there are some modifications due to the new weighting scheme. They are given as follows: 
	\begin{eqnarray}
	\label{beta0mmllts1} \hat{\beta}_{0} = \bar{y}_{[\cdot]}-\overline{\mathbf{x} }'_{[\cdot]}\boldsymbol{\hat\beta}+ \dfrac{\Delta}{w} \hat{\sigma}
	\end{eqnarray}
	where 
	\begin{eqnarray*}
		\bar{y}_{[\cdot]}= \frac{\displaystyle
			\sum_{i=1}^n \delta_i \delta_i^x y_{i}}{w}, 
		\quad \bar{\mathbf{x}}_{[\cdot] j}= \frac{\displaystyle\sum_{i=1}^n \delta_i \delta_i^x x_{ij}}{w},   \quad \overline{\mathbf{x} }'_{[\cdot]}&=&[\bar{\mathbf{x}}_{[\cdot]1} \quad  \bar{\mathbf{x}}_{[\cdot]2} \quad  \cdots \quad  \bar{\mathbf{x}}_{[\cdot] m} ],
	\end{eqnarray*}
	\begin{eqnarray*}
		w= \displaystyle\sum_{i=1}^n \delta_i  \delta_i^x, \quad  \Delta = \displaystyle \sum_{i=1}^n \alpha_i \delta_i^x. 
	\end{eqnarray*}	
	\begin{eqnarray*}
		\boldsymbol{\delta}= diag(\delta_i \delta_i^x) \quad \textrm{ and } \quad  \boldsymbol{\alpha}= diag(\alpha_i \delta_i^x), \quad i=1,2,\ldots,n. 
	\end{eqnarray*}
	The remaining expressions, i.e.~$\hat{\boldsymbol{\beta}}$, $\hat{\sigma}$, $\mathbf{K}$, $\mathbf{L}$, $B$ and $C$ are formulated in the same way as given in the previous section, but using the weights and formulations as introduced above. 
	
\end{description}

It should be noticed that RAMML estimators are defined similar to generalized $M$ (GM) estimators which are robust to both types of outliers, see e.g. Maronna et al.  \cite{Maronnaetal2006} for further details. 

\subsection{Computation of RAMML estimators}
The RAMML estimators are obtained after two iterations. In the first iteration, an initial estimate of $t_i$ is required. Unlike Donmez \cite{Donmez2010}, we do not consider a reparametrization of model \eqref{eq:linearmodel}. We continue with model \eqref{eq:linearmodel} and  suggest to use well-known robust regression estimators such as LTS 
or S to initialize the iterations. In other words, the initial estimate of $t_{i}$ is proposed as 
\begin{equation}\label{eq:adaptivemmlt5}
t_{i}=\frac{y_{i}-T_{0}-\displaystyle\sum_{j=1}^n x_{ij}T_{j}}{S_0} ,
\end{equation}
where $T_0$, $T_j$ ($j=1,2,\ldots,m$) and $S_0$ are  obtained from  LTS or S regression.   After getting the initial estimate of $t_i$,  initial RAMML estimators denoted by $\tilde{\beta}_{0}$, $\tilde{\beta}_{j}$ and $\tilde{\sigma}$ are obtained following Steps 1, 2* and 3*.

In the second iteration,  $t_i$  is updated with the initial RAMML estimators obtained in the previous iteration as follows: 
\begin{equation}
t_{i}=\frac{y_{i}-\tilde{\beta}_{0}-\displaystyle\sum_{j=1}^n x_{ij}\tilde{\beta}_{j}}{\tilde{\sigma}}
\end{equation}
and Steps 1, 2* and 3* are utilized. The final RAMML estimators are obtained at the end of  Step 3* using Equations \eqref{betammllts}, \eqref{sigmammllts} and \eqref{beta0mmllts1}. 

The resulting estimators obtained using LTS and S regression are denoted by  RAMML$_1$ and RAMML$_2$, respectively.  It is clear that RAMML estimators are also explicitly formulated and they can be computed after  two simple iterations. Therefore, they are computationally straightforward. 

If the weights $\delta_i^x$ are removed from the estimating equations, RAMML$_1$ and RAMML$_2$  will reduce to the AMML$_1$ and AMML$_2$, respectively.  It should be noted that these AMML$_1$ and AMML$_2$ estimators differ from that of Donmez (2010) since the inital $t_i$s are defined differently.

It should be noted that different initial estimates can be preferred for finding initial $t_i$s. The important issue here is that the initial estimators should be robust so that the resulting estimator is robust.  In this study, we consider two well-known robust estimators as initials, i.e.~LTS and S. We also consider the LMS estimator as an initial estimator and obtain more or less the same results, see Acitas et al.  \cite{Acitasetal2019}. However, we do not consider it here because it is not computationally efficient.

\section{Simulation study}\label{sec:simulation}
In this section, we conduct a simulation study to compare the performances of the proposed estimators RAMML$_1$ and RAMML$_2$  along with some known robust estimators such as MM, LTS and S estimators under different simulation scenarios. AMML$_1$ and AMML$_2$ are also taken into account during the performance comparisons. 

The simulation setup and contamination schemes are considered similar to those used by Gschwandtner and Filzmoser \cite{GschwandtnerFilzmoser2012}.  Model \eqref{eq:linearmodel} is used during the simulation study where $\beta_0$ is assumed to be 0 without loss of generality. The data are generated as follows:
\begin{eqnarray*}
	\mathbf{x}_i &\sim & \mathbf{N}_m(\mathbf{0}, \mathbf{I}_m), \quad i=1,2,\ldots,n \\
	\beta_j &=& 1/\sqrt{m},  \quad j=1,2,\ldots,m \quad \text{and thus} \quad \norm{\boldsymbol{\beta}}=1.
\end{eqnarray*}
Here, $\mathbf{N}_m$ denotes the $m-$dimesional multivariate normal distribution, $\mathbf{0}$ stands for the vector of zeros and $\mathbf{I}_m$ is the identity matrix.

Five different error distributions are considered in model \eqref{eq:linearmodel}: (i) $N(0,1)$, (ii) Laplace, (iii) $t_5$, (iv) $t_1$ (Cauchy) and (v) slash distribution.  The leverage points are generated replacing the first $n_{out}$ observations with the following versions:
\begin{equation*}
\mathbf{x}_i =[\begin{matrix}
\ell & \ell & \cdots &\ell \\
\end{matrix}]'  \quad \text{and} \quad  y_i=\mathbf{x}_i '\mathbf{a} \quad \text{for} \quad i=1,2,\ldots,n
\end{equation*}
where $\mathbf{a}$ is a normalized unit vector defined by $\mathbf{a}=\boldsymbol{\nu}- (\boldsymbol{\nu}'\boldsymbol{\beta})\boldsymbol{\beta}.$
Here, $\boldsymbol{\nu}$ is a vector having entries $(-1)^j$.  If $m=1$, $a$ is taken as -1.  We consider two different versions for leverage points, i.e. $\ell=5$ and $\ell=10$. This is because of the fact that we would like to see how  the performances of the estimators  are affected as the magnitude of the leverage point increases. The sample size is taken to be $n=50$ and $200$. The number of the predictors $m$ taken from the set $\{1,5,10,20\}$.  $n_{out}$ is the number of leverage points and determined according to two contamination levels, i.e. 10\% and 20\%.

The mean squared error (MSE) formulated as 
\begin{equation}\label{eq:mse}
MSE=\frac{1}{n_{rep}} \sum_{r=1}^{n_{rep}} \norm{{\hat{\boldsymbol{\beta}}}^r-\boldsymbol{\beta}}^2
\end{equation} 
is calculated to evaluate the performances of the estimators based on $n_{rep}=500$ Monte-Carlo runs. All computations are carried out using the software environment R. The results are tabulated in Tables \ref{tab:simres1} and  \ref{tab:simres2}.  It should be mentioned that LS estimators are not considered in these tables since they are not robust to outliers. 

\begin{center}
	\textbf{[Table \ref{tab:simres1} is here ]}
\end{center}

It can be seen from Table \ref{tab:simres1} that the MSEs of the RAMML$_1$ and RAMML$_2$ are close to each other and smaller than those of MM, LTS, AMML$_1$, S and AMML$_2$ for both 10\% and 20\% contamination levels when the error distribution is normal, Laplace, $t_5$, $t_1$ and slash.  The RAMML$_2$ estimator is almost the best since it has the 
smallest MSE. However, the RAMML$_1$ also performs better than the others in some cases, 
i.e. if the contamination level is 10\%, $m=1$, 5 and 10, when the error distribution is normal.  It is also the best estimator for simple linear regression with normal, Laplace and $t_5$ error distributions when the contamination level is 20\%. In the remaining cases, the  RAMML$_2$ is preferable to other estimators. It should also be mentioned that the  AMML$_2$  performs better than  MM, LTS, AMML$_1$, RAMML$_1$,  S and RAMML$_2$ when the number of predictors is 20, $\ell = 5$ and the contamination level is   20\%. The MM estimator has a remarkable performance but it is not sufficient to be the best. In short, the RAMML$_2$ is slightly better than its 
competitors in most cases considered in Table  \ref{tab:simres1}.

\begin{center}
	\textbf{[Table \ref{tab:simres2} is here ]}
\end{center}

When the sample size is increased to 200, see Table~\ref{tab:simres2}, we 
can draw similar conclusions to those obtained from  Table  \ref{tab:simres1}. However, here the MM estimator gains efficiency and performs better than the other estimators for some cases. For example, for simple linear regression with normal, Laplace and $t_5$ error distributions for 10\% level of contamination and $\ell = 5$, it has the minimum MSE value, 
followed by the RAMML$_2$. If the distribution of the error is heavy-tailed, i.e.~$t_1$ and slash, the RAMML$_2$ is the best. This conclusion is also true for $\ell = 5$. If $m$ is increased from 5 to 20,  the RAMML$_2$ is mostly the best estimator when the contamination level is 10\%. In some cases such as $\ell =5$, $m=10$, $t_1$ error distribution and $\ell =10$, $m=20$, normal distribution, the RAMML$_1$ outperforms the other estimators. When the level of contamination is 20\%, the RAMML$_1$ and RAMML$_2$ outperform the others. For the simple linear regression case with all error distributions, the RAMML$_2$ is better than MM, LTS, AMML$_1$, RAMML$_1$, S and AMML$_2$. For $m\in \{5,10,20\}$, the RAMML$_1$ is more preferable for normal, Laplace and $t_5$ error distributions. However, if the error distribution is heavy-tailed, RAMML$_2$ is the best regardless of the value of $\ell$ and contamination level. It can be concluded that  RAMML$_2$  is the best estimator for almost all cases considered in Table \ref{tab:simres2}.

In this part of the simulation study, we just focus on the MSEs of the estimators for the scale parameter $\sigma$ when the distribution of the error terms is $N(0,1)$.  The MSE of $\hat{\sigma}$ is calculated using \eqref{eq:mse} in which the symbols $\boldsymbol{\beta}$ are replaced by $\sigma$. The results are tabulated in Table \ref{tab:simres3}. The MSEs of RAMML$_1$ and  RAMML$_2$  are more or less the same.  RAMML$_1$ and/or   RAMML$_2$ has the minimum MSE value in most of the cases. 

\begin{center}
	\textbf{[Table \ref{tab:simres3} is here ]}
\end{center}

It is well-known that the trade-off between robustness and efficiency is important. We therefore explore the efficiencies of RAMML$_1$ and RAMML$_2$ under normality in this part of the simulation study. For this purpose, the performances of all estimators considered in this study are compared  with the LS estimators  when the error distribution is $N(0,1)$ with zero contamination.  The results are given  in Table~\ref{tab:simres4}. It is clear from this table that the MSE values of  RAMML$_1$ and RAMML$_2$  are close to those of LS. This conclusion implies that RAMML$_1$ and RAMML$_2$ are not only robust to both types of outliers but also efficient under normality. 

\begin{center}
	\textbf{[Table \ref{tab:simres4} is here ]}
\end{center}

Finally, we explore the performances of the estimators when the predictors are correlated. Therefore, we make a simple modification in the simulation scenario given at the beginning of this section. Here,  the  $\mathbf{x}_i $s are generated from $\mathbf{N}_m(\mathbf{0}, \mathbf{V})$ where $\mathbf{V}=[v_{jk}]$ with $v_{jj}=1$ and $v_{jk}=\rho$ for $j\neq k$ and $k=1,2,\ldots,m$.   In Table  \ref{tab:simres.correlated}, we report the MSEs of the estimators for  $\rho=0.90$, $\ell=10$ and contamination levels 10\%.   This table 
shows that the MM estimator is the best one in almost all settings, mostly followed by RAMML$_1$ and RAMML$_2$.

Overall, the MSEs of RAMML$_1$ and  RAMML$_2$ are close to each other and  they outperform their competitors when there are $x-$outliers.  Among these two estimators, we see that RAMML$_2$  is preferable.

\begin{center}
	\textbf{[Table \ref{tab:simres.correlated} is here ]}
\end{center}

\section{Application}
In this part of the study, we provide two data sets which are widely used in the literature to show the implementation of the proposed estimators. The full data sets can be found in the R package ``\texttt{robustbase}''.

\subsection{Hertzsprung-Russell Diagram Data of Star Cluster CYG OB1}

\begin{figure}[htp]
	\begin{center}
		\includegraphics[scale=0.6]{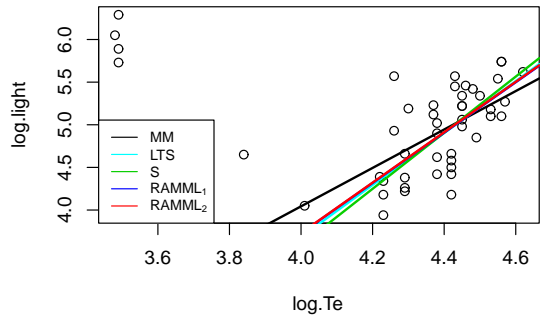}
	\end{center}
	\caption{Fitted regression lines to the starsCYG data.} \label{fig:starsCYG}
\end{figure}

This data set is shortly denoted as starsCYC and includes 47 observations with two variables named log.Te ($x$) and log.light ($y$). The former variable is  the logarithm of the effective temperature at the surface of the star  while the latter one is the logarithm of its light intensity, see Rousseeuw and Leroy \cite{RousseeuwLeroy1987}. 
It is clear from Figure \ref{fig:starsCYG} that the starsCYC data contains outliers, i.e.  four stars in the left upper corner are far away from bulk of the data, and thus they are leverage points.  Rousseeuw and Leroy \cite{RousseeuwLeroy1987} employ the LMS method and obtain more reliable results than for the LS method.  Different from them, in this study, we use the RAMML$_1$ and RAMML$_2$ estimation methods along with the other methods considered in the previous section  to estimate the regression coefficients.  To evaluate the performances of the methods, we use the standard error of prediction (SEP)  criterion  formulated as
\begin{equation*}
SEP=\sqrt{\dfrac{1}{n-1} \sum_{i=1}^n (y_i-\hat{y}_i-\text{bias})^2}, \quad \text{bias}=\dfrac{1}{n} \sum_{i=1}^n (y_i-\hat{y}_i)
\end{equation*}
where, $\hat{y}_i$ denotes the fitted value for the $i$-th observation. It should be noted that using  SEP  may give misleading results in case of outliers since each observation has the same contribution in the formula of SEP. Therefore, a robust version of the SEP is required.  We therefore compute the trimmed version of SEP where the trimming value is 10\%.  The robust performance criterion will lead to a contrary picture since the influence of the  largest contributions is reduced. There is no general rule for the trimming value and thus it is subjective. One can take lower or higher values according to the data quality, see for example Liebmann et al. \cite{Liebmannetal2010}, Filzmoser and Todorov \cite{Filzmoseretal2011}.

\begin{center}
	\textbf{[Table \ref{tab:starCYGestimates} is here ]}
\end{center}

The estimated values of the regression coefficients and the scale parameter and also SEP and SEP$_{\text{trim}}$  values  are tabulated in Table~\ref{tab:starCYGestimates}.  The SEP$_{\text{trim}}$ values suggest that RAMML$_1$ and RAMML$_2$  are slightly preferable to MM, LTS, AMML$_1$, S and AMML$_2$.  Furthermore, the scale estimates obtained based on RAMML$_1$ and RAMML$_2$ are much smaller  than those of other estimation methods. This conclusion implies that estimates obtained by RAMML$_1$ and RAMML$_2$ give more reliable results. The scatter plot of the starsCYG data and the fitted regression lines are also illustrated in Figure \ref{fig:starsCYG}.  The fitted regression lines obtained from the RAMML$_1$ and RAMML$_2$ estimates are also satisfactory.  

\begin{figure}[htp]
	\begin{center}
		\includegraphics[scale=0.6]{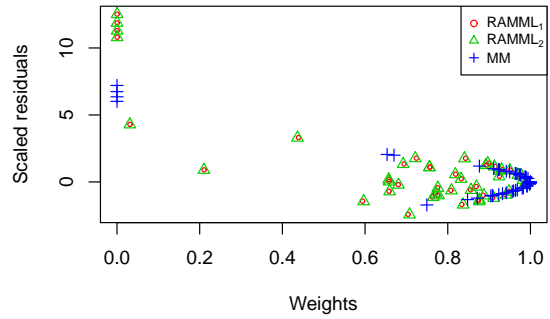}
	\end{center}
	\caption{Weights versus scaled residuals for the estimators MM, RAMML$_1$ and RAMML$_2$, based
		on the starsCYG data.} \label{fig:starsCYGwvsres}
\end{figure}

The weights obtained at the final step of the robust estimation procedures MM, RAMML$_1$ and RAMML$_2$ are plotted against the scaled residuals in Figure~\ref{fig:starsCYGwvsres}. The weights obtained from RAMML$_1$ and RAMML$_2$ are smaller than those of MM. This is because the weights given by the RAMML methodology allow more flexiblity to the data compared to those 
from the MM, see also Figure~\ref{fig:starsCYG}.

%
\begin{center}
	\textbf{[Figure  \ref{fig:starsCYG} is here ]}
\end{center}

\begin{center}
	\textbf{[Figure  \ref{fig:starsCYGwvsres} is here ]}
\end{center}

\subsection{Aircraft data}
This data set consisting of 23 obervations on single-engine aircrafts is obtained from 
the Office of Naval Research over the years 1947-1979, see Gray \cite{Gray1985} and  Rousseeuw and Leroy \cite{RousseeuwLeroy1987}. 
It contains four predictors and a response variable whose descriptions are given below: 
\begin{eqnarray*}
	y=\text{cost (in units of \$100,000)}, \quad 	x_1=\text{aspect ratio}, \quad 	x_2 =\text{lift-to-drag ratio},
\end{eqnarray*}
\begin{eqnarray*}
	x_3=\text{weight of plane (in pounds)}\quad \text{and} \quad 	x_4 =\text{maximal thrust.}
\end{eqnarray*}

We use the methods MM, LTS, AMML$_1$, RAMML$_1$, S,  AMML$_2$, and RAMML$_2$ to estimate the unknown parameters.  Robust diagnostics obtained based on LTS show that there are two outliers in the data. The estimated values of the regression coefficients and the scale parameter along with performance criteria are shown in Table~\ref{tab:aircraft}. The minimum SEP$_{\text{trim}}$  value is obtained for LTS which is followed by MM, RAMML$_1$ and RAMML$_2$. However, the scale estimates obtained from RAMML$_1$ and RAMML$_2$ are smaller than those of LTS and MM. Therefore, RAMML$_1$ and RAMML$_2$ are also promising and can be considered as good alternative estimators.

\begin{figure}[htp]
	\begin{center}
		\includegraphics[scale=0.6]{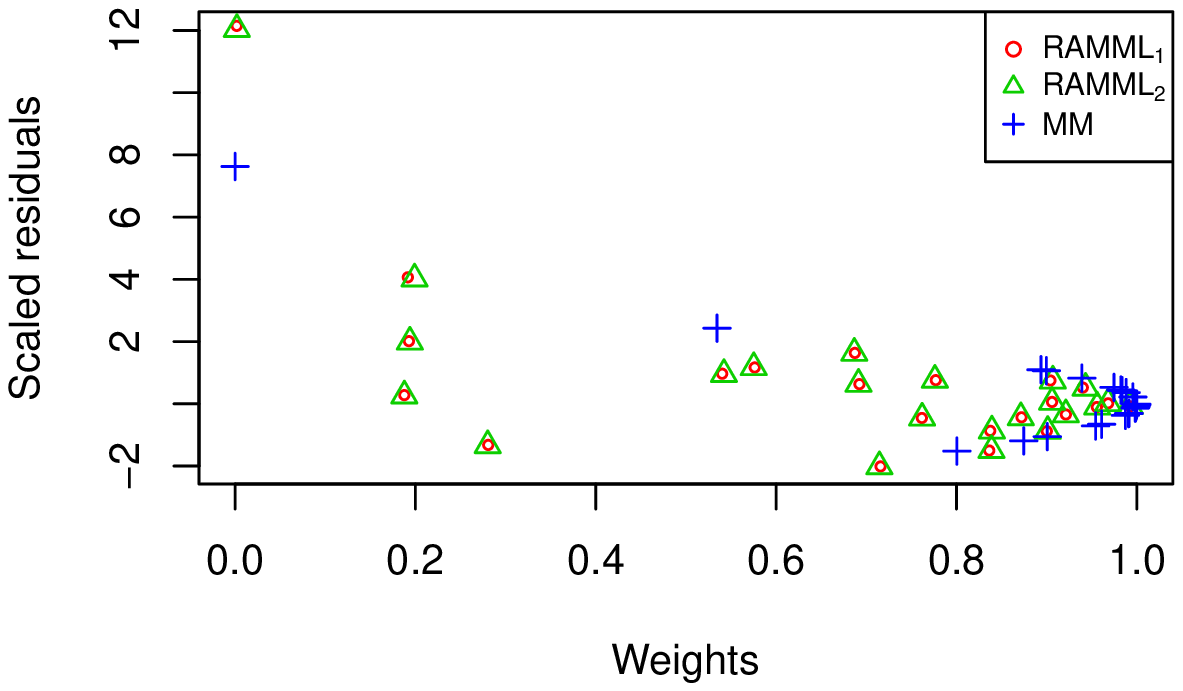}
	\end{center}
	\caption{Weights versus scaled residuals for the Aircraft data.} \label{fig:aircraftwvsres}
\end{figure}

In Figure \ref{fig:aircraftwvsres}, the weights obtained at the last  step of the robust estimation procedures MM, RAMML$_1$ and RAMML$_2$ are plotted aganist the scaled residuals.  Also in this example, the RAMML weights are  smaller 
than those of the MM estimator; this is because the RAMML weights are more flexible.
\begin{center}
	\textbf{[Table \ref{tab:aircraft} is here ]}
\end{center}

\begin{center}
	\textbf{[Figure  \ref{fig:aircraftwvsres} is here ]}
\end{center}

\section{Conclusions}

In this study, we propose RAMML estimators which are robust to both  $y$ and $x$ outliers. The RAMML estimators are obtained  by downweighting $x-$outliers using an extra weight function in the AMML estimation method.  

The RAMML estimators have some advantage over existing robust estimators: 
they  are (i)  explicitly formulated and thus easy to compute, (ii) efficient  under the normality assumption and (iii) more preferable according to the MSE and SEP criteria.  These properties make them good alternatives to existing robust estimators. 
The simulation studies have also shown that the RAMML estimators are improving over the other considered robust
estimators if the contamination level increases, in case of heavy-tailed error distributions, and if the 
number of explanatory variables increases. Thus, in more difficult data situations there the advantage gets
more pronounced. Another interesting feature of the RAMML estimators is that the estimated residual variance
is in many situation smaller than for the alternatives. Together with the other properties, this is a useful feature
which is based on a more flexible weighting scheme within the computation of the estimator.

Although the new robust estimators show good performance, there are still some points that should be considered. One of them is the choice of the weights $\delta_i^x$ assigned to the predictors.   
Different choices of $\delta_i^x$ will affects the efficiencies of the RAMML estimators. 
In this study, the weights $\delta_i^x$ are obtained  similar to  the weights $\delta_i$. In the future 
studies, alternative weight functions such as Huber, Fair, etc. will be considered.  

\nolinenumbers
\small

\section*{Acknowledgements}
This study is supported by ``The Scientific and Technological Research Council of Turkey (TUBITAK)'' as part of the ``2219 -- International Postdoctoral Research Scholarship Programme''.

\newpage

\begin{table}[h]
	\caption{The simulated MSEs of MM, LMS, AMML$_1$, RAMML$_1$,   LTS, AMML$_2$, RAMML$_2$ estimators for regression coefficients: $n =50$.} \label{tab:simres1}
	\begin{center}
		\scriptsize{\begin{tabular}{l ccccc c ccccc}
				\hline
				Estimator	&	Normal	&	Laplace	&	$t_5$	&	$t_1$&	Slash	&		&	Normal	&	Laplace	&	$t_5$	&	$t_1$	&	Slash	\\
				\hline	\hline
				\multicolumn{12}{c}{Contamination level = $10\%$} \\
				\hline
				&	\multicolumn{11}{c}{$m=1$} \\
				\cline{2-12} 
				&	\multicolumn{5}{c}{$\ell=5$}   &	 & 	\multicolumn{5}{c}{ $\ell=10$} 			\\
				\cline{2-6}  \cline{8-12}
				MM	&	0.0474	&	0.0831	&	0.0519	&	1.1207	&	2.2697	&	&	0.0397	&	0.0774	&	0.0422	&	0.4349	&	1.4712	\\
				
				LTS	&	0.1111	&	0.1423	&	0.1665	&	0.3463	&	1.2311	&	&	0.1181	&	0.1363	&	0.1534	&	0.3616	&	1.2682	\\
				
				AMML$_1$	&	0.8011	&	1.2815	&	1.1698	&	2.1221	&	2.2956	&	&	0.2737	&	0.8501	&	0.6461	&	2.6837	&	3.3277	\\
				
				RAMML$_1$	&	0.0323	&	0.0476	&	0.0437	&	0.1808	&	0.3142	&	&	0.0295	&	0.0419	&	0.0355	&	0.1640	&	0.2536	\\
				
				S	&	0.0856	&	0.1003	&	0.0803	&	0.3766	&	1.7704	&	&	0.0800	&	0.0902	&	0.0702	&	0.4027	&	1.5348	\\
				
				AMML$_2$	&	0.6775	&	1.0634	&	1.0226	&	2.0572	&	2.2894	&	&	0.1907	&	0.5905	&	0.4307	&	2.4178	&	3.2486	\\
				
				RAMML$_2$	&	0.0321	&	0.0463	&	0.0430	&	0.1687	&	0.3048	&	&	0.0296	&	0.0415	&	0.0354	&	0.1556	&	0.2419	\\
				\hline
				&	\multicolumn{11}{c}{$m=5$} \\
				\cline{2-12} 
				MM	&	0.6763	&	0.8421	&	0.7994	&	1.3103	&	1.8802	&	&	0.7433	&	0.8264	&	0.8419	&	1.3939	&	1.8336	\\
				
				LTS	&	1.1991	&	1.2970	&	1.3264	&	1.6641	&	2.4065	&	&	1.1903	&	1.3648	&	1.3543	&	1.7104	&	2.4053	\\
				
				AMML$_1$	&	0.9573	&	1.1174	&	1.0607	&	1.5667	&	2.1197	&	&	0.9339	&	1.0755	&	1.0617	&	1.6799	&	2.1338	\\
				
				RAMML$_1$	&	0.3569	&	0.4778	&	0.4295	&	0.9849	&	1.5785	&	&	0.1352	&	0.2245	&	0.1906	&	0.7523	&	1.2911	\\
				
				S	&	0.9484	&	1.2203	&	1.2053	&	1.7732	&	2.9104	&	&	1.0298	&	1.1326	&	1.2092	&	1.8727	&	2.6679	\\
				
				AMML$_2$	&	0.8496	&	1.0715	&	0.9965	&	1.5387	&	2.0872	&	&	0.7432	&	0.9161	&	0.8933	&	1.6366	&	2.0738	\\
				
				RAMML$_2$	&	0.2978	&	0.4413	&	0.3946	&	0.9560	&	1.5511	&	&	0.1367	&	0.2235	&	0.1961	&	0.7183	&	1.2272	\\				
				\hline
				&	\multicolumn{11}{c}{$m=10$} \\
				\cline{2-12} 
				MM	&	1.2647	&	1.6644	&	1.5661	&	2.5889	&	3.798	&	&	1.2768	&	1.6596	&	1.6618	&	2.7727	&	3.6821	\\
				
				LTS	&	2.5168	&	3.1116	&	2.9643	&	4.1231	&	7.4949	&	&	2.4722	&	3.0758	&	2.9561	&	4.0753	&	6.1214	\\
				
				AMML$_1$	&	1.3476	&	1.6566	&	1.5553	&	2.9153	&	4.5046	&	&	1.3567	&	1.6161	&	1.5779	&	3.1962	&	4.1157	\\
				
				RAMML$_1$	&	0.9400	&	1.218	&	1.1349	&	2.4783	&	4.0340	&	&	0.3106	&	0.5558	&	0.4720	&	2.0974	&	3.1041	\\
				
				S	&	2.0672	&	2.8438	&	2.6515	&	4.3687	&	6.9140	&	&	2.1380	&	2.7441	&	2.6572	&	4.1898	&	6.5537	\\
				
				AMML$_2$	&	1.1733	&	1.5749	&	1.4535	&	2.8347	&	4.2468	&	&	1.0396	&	1.4531	&	1.3917	&	3.0774	&	3.9912	\\
				
				RAMML$_2$	&	0.7825	&	1.1355	&	1.0536	&	2.4038	&	3.7700	&	&	0.3146	&	0.5553	&	0.4759	&	1.9827	&	2.9669	\\
				\hline
				&	\multicolumn{11}{c}{$m=20$} \\
				\cline{2-12} 
				MM	&	3.8055	&	5.4072	&	5.1708	&	11.4061	&	19.3850	&	&	3.7672	&	5.3245	&	5.1794	&	12.7201	&	19.0190	\\
				
				LTS	&	7.1310	&	9.4504	&	8.9798	&	20.3109	&	35.0990	&	&	7.5826	&	9.4996	&	9.6198	&	21.3224	&	33.3410	\\
				
				AMML$_1$	&	2.8865	&	3.7237	&	3.4436	&	9.6187	&	16.4300	&	&	3.0323	&	3.9743	&	3.6027	&	9.9341	&	16.4600	\\
				
				RAMML$_1$	&	2.7306	&	3.6004	&	3.2992	&	9.3477	&	16.1550	&	&	1.2295	&	2.3787	&	1.9306	&	8.6733	&	15.1420	\\
				
				S	&	6.0778	&	8.2255	&	7.5602	&	20.5418	&	35.2120	&	&	5.9019	&	8.7704	&	7.7262	&	20.0038	&	29.1150	\\
				
				AMML$_2$	&	2.4717	&	3.3454	&	3.1723	&	10.0115	&	15.4510	&	&	2.4700	&	3.4223	&	3.0899	&	9.6530	&	15.5550	\\
				
				RAMML$_2$	&	2.2657	&	3.1453	&	2.9610	&	9.6906	&	15.0810	&	&	0.9957	&	1.9028	&	1.5561	&	8.3624	&	14.3660	\\
				\hline
				\hline
				\multicolumn{12}{c}{Contamination level = $20\%$} \\
				\hline
				&	\multicolumn{11}{c}{$m=1$} \\
				\cline{2-12} 
				MM	&	0.9063	&	1.5115	&	1.3348	&	2.8098	&	3.1718	&	&	1.1198	&	1.6951	&	1.6449	&	3.1743	&	3.6786	\\
				
				LTS	&	1.9771	&	2.1900	&	2.1504	&	2.9506	&	3.4027	&	&	2.2838	&	2.3322	&	2.6162	&	3.1222	&	3.7683	\\
				
				AMML$_1$	&	2.6602	&	2.8145	&	2.7665	&	2.9762	&	3.0118	&	&	2.6405	&	3.1454	&	3.1654	&	3.6689	&	3.6903	\\
				
				RAMML$_1$	&	0.2286	&	0.3897	&	0.3260	&	0.7351	&	0.9377	&	&	0.0328	&	0.0530	&	0.0481	&	0.1758	&	0.2441	\\
				
				S	&	1.0514	&	1.7672	&	1.5666	&	3.1428	&	3.6572	&	&	1.1219	&	1.7432	&	1.7081	&	3.2864	&	3.8538	\\
				
				AMML$_2$	&	2.4319	&	2.7249	&	2.6339	&	2.9805	&	3.0240	&	&	1.8558	&	2.7676	&	2.6542	&	3.6578	&	3.6942	\\
				
				RAMML$_2$	&	0.1452	&	0.3293	&	0.2638	&	0.7275	&	0.9437	&	&	0.0333	&	0.0532	&	0.0488	&	0.1690	&	0.2341	\\
				\hline
				&	\multicolumn{11}{c}{$m=5$} \\
				\cline{2-12} 
				MM	&	1.3353	&	1.4435	&	1.3949	&	1.7396	&	2.0893	&	&	1.3779	&	1.4804	&	1.4346	&	1.6947	&	2.0291	\\
				
				LTS	&	2.3538	&	2.6105	&	2.4865	&	3.2080	&	4.4527	&	&	2.4167	&	2.6958	&	2.5875	&	3.2793	&	4.4298	\\
				
				AMML$_1$	&	1.2048	&	1.2656	&	1.2523	&	1.7383	&	2.1198	&	&	1.2499	&	1.3365	&	1.2883	&	1.7571	&	2.1580	\\
				
				RAMML$_1$	&	0.9746	&	1.0474	&	1.0370	&	1.5205	&	1.8879	&	&	0.1904	&	0.3055	&	0.2579	&	0.8240	&	1.3294	\\
				
				S	&	2.1677	&	2.4305	&	2.3244	&	2.9160	&	3.8286	&	&	2.2236	&	2.4929	&	2.3276	&	2.8058	&	3.8126	\\
				
				AMML$_2$	&	1.1977	&	1.2625	&	1.2452	&	1.7209	&	2.0833	&	&	1.2289	&	1.3164	&	1.2828	&	1.7381	&	2.1254	\\
				
				RAMML$_2$	&	0.9624	&	1.0412	&	1.0248	&	1.5008	&	1.8485	&	&	0.1847	&	0.2953	&	0.2536	&	0.8018	&	1.2893	\\
				\hline
				&	\multicolumn{11}{c}{$m=10$} \\
				\cline{2-12}
				MM	&	2.7202	&	3.3625	&	3.3854	&	4.6944	&	6.0067	&	&	2.9763	&	3.4669	&	3.1312	&	4.4531	&	5.9192	\\
				
				LTS	&	4.3832	&	5.7078	&	5.3309	&	8.7250	&	12.4536	&	&	4.3648	&	5.6491	&	5.2871	&	9.4753	&	14.6904	\\
				
				AMML$_1$	&	1.8994	&	2.2494	&	2.1421	&	3.6946	&	5.0060	&	&	1.9994	&	2.2793	&	2.2528	&	3.8856	&	5.301	\\
				
				RAMML$_1$	&	1.8389	&	2.1956	&	2.0839	&	3.6169	&	4.9443	&	&	0.9558	&	1.3027	&	1.2283	&	3.0656	&	4.6646	\\
				
				S	&	3.8241	&	4.7879	&	4.6691	&	7.5793	&	10.5919	&	&	4.1517	&	5.1859	&	4.5923	&	8.2151	&	11.6736	\\
				
				AMML$_2$	&	1.6618	&	1.9274	&	1.8949	&	3.3243	&	4.5066	&	&	1.7220	&	2.0002	&	1.9096	&	3.3479	&	4.4427	\\
				
				RAMML$_2$	&	1.5641	&	1.8257	&	1.7968	&	3.1857	&	4.3612	&	&	0.6103	&	0.9395	&	0.7995	&	2.4694	&	3.7191	\\
				\hline
				&	\multicolumn{11}{c}{$m=20$} \\
				\cline{2-12}  
				MM	&	10.6851	&	13.8939	&	13.6416	&	50.8300	&	87.5050	&	&	11.3025	&	14.3318	&	13.6001	&	52.1760	&	77.6950	\\
				
				LTS	&	13.5703	&	19.2541	&	19.0750	&	89.8350	&	126.6330	&	&	15.2065	&	20.6194	&	18.4261	&	83.5840	&	123.8700	\\
				
				AMML$_1$	&	12.7179	&	17.6868	&	17.7259	&	80.4410	&	112.0310	&	&	14.2021	&	19.1793	&	17.0929	&	74.2650	&	107.8330	\\
				
				RAMML$_1$	&	12.7966	&	17.8217	&	17.8598	&	81.1750	&	113.1480	&	&	14.2039	&	19.2348	&	17.1026	&	74.7500	&	108.9300	\\
				
				S	&	14.2006	&	20.8200	&	18.3638	&	143.607	&	143.8900	&	&	13.5973	&	19.8771	&	18.3664	&	95.4700	&	115.5300	\\
				
				AMML$_2$	&	5.0160	&	7.5507	&	5.4790	&	53.6650	&	40.8890	&	&	5.5229	&	7.1065	&	7.4644	&	32.7620	&	41.8120	\\
				
				RAMML$_2$	&	5.1211	&	7.7963	&	5.6209	&	54.7630	&	42.1430	&	&	4.9299	&	6.6159	&	6.9766	&	33.1670	&	42.4460	\\
				\hline
		\end{tabular}}
	\end{center}
\end{table}

\newpage
\begin{table}[h]
	\caption{The simulated MSEs of MM, LMS, AMML$_1$, RAMML$_1$,   LTS, AMML$_2$, RAMML$_2$ estimators for regression coefficients: $n=200$.} \label{tab:simres2}
	\begin{center}
		\scriptsize{\begin{tabular}{l ccccc c ccccc}
				\hline
				Estimator	&	Normal	&	Laplace	&$t_5$	&	$t_1$&	Slash	&		&	Normal	&	Laplace	&$t_5$	&	$t_1$	&	Slash	\\
				\hline\hline
				\multicolumn{12}{c}{Contamination level = $10\%$} \\
				\hline
				&	\multicolumn{11}{c}{$m=1$} \\
				\cline{2-12} 
				&	\multicolumn{5}{c}{  $\ell=5$}   &	 & 	\multicolumn{5}{c}{  $\ell=10$} 			\\
				\cline{2-6}  \cline{8-12}
				MM	&	0.0060	&	0.0085	&	0.0078	&	0.5136	&	2.4011	&	&	0.3472	&	0.4862	&	0.5819	&	1.1543	&	1.3631	\\
				
				LTS	&	0.0069	&	0.0087	&	0.0090	&	0.0257	&	0.9422	&	&	0.8945	&	0.7963	&	1.0423	&	1.2114	&	1.5357	\\
				
				AMML$_1$	&	0.6964	&	1.2441	&	1.1109	&	2.1133	&	2.2443	&	&	0.9002	&	1.0152	&	1.0219	&	1.2870	&	1.4867	\\
				
				RAMML$_1$	&	0.0071	&	0.0128	&	0.0100	&	0.0561	&	0.1066	&	&	0.4557	&	0.4614	&	0.5554	&	0.9008	&	1.1409	\\
				
				S	&	0.0147	&	0.0095	&	0.0156	&	0.0327	&	1.1916	&	&	0.5249	&	0.6682	&	0.8161	&	1.4541	&	1.8625	\\
				
				AMML$_2$	&	0.5764	&	0.9927	&	0.9086	&	2.0420	&	2.2402	&	&	0.6607	&	0.9147	&	0.8968	&	1.2699	&	1.4605	\\
				
				RAMML$_2$	&	0.0071	&	0.0122	&	0.0098	&	0.0493	&	0.1023	&	&	0.2258	&	0.3525	&	0.3854	&	0.9031	&	1.1201	\\
				\hline
				&	\multicolumn{11}{c}{$m=5$} \\
				\cline{2-12} 
				MM	&	0.2626	&	0.3484	&	0.4125	&	0.9128	&	1.0750	&	&	0.2958	&	0.3375	&	0.4387	&	0.9928	&	1.1468	\\
				
				LTS	&	0.7012	&	0.5478	&	0.7708	&	0.8773	&	1.1372	&	&	0.7343	&	0.5465	&	0.8001	&	0.9557	&	1.2046	\\
				
				AMML$_1$	&	0.8465	&	0.9088	&	0.9050	&	1.0321	&	1.1099	&	&	0.7398	&	0.8282	&	0.8656	&	1.1169	&	1.2055	\\
				
				RAMML$_1$	&	0.1948	&	0.2184	&	0.2537	&	0.4840	&	0.6080	&	&	0.0296	&	0.0467	&	0.0407	&	0.1427	&	0.2510	\\
				
				S	&	0.2876	&	0.3808	&	0.4624	&	1.0463	&	1.2826	&	&	0.3312	&	0.3912	&	0.4801	&	1.0471	&	1.3217	\\
				
				AMML$_2$	&	0.7168	&	0.8596	&	0.8426	&	1.0253	&	1.1002	&	&	0.4003	&	0.6415	&	0.6182	&	1.1031	&	1.1935	\\
				
				RAMML$_2$	&	0.0900	&	0.1616	&	0.1655	&	0.4853	&	0.6009	&	&	0.0298	&	0.0464	&	0.0409	&	0.1329	&	0.2359	\\
				\hline
				&	\multicolumn{11}{c}{$m=10$} \\
				\cline{2-12} 
				MM	&	0.3472	&	0.4862	&	0.5819	&	1.1543	&	1.3631	&	&	0.4094	&	0.5243	&	0.5526	&	1.1824	&	1.4381	\\
				
				LTS	&	0.8945	&	0.7963	&	1.0423	&	1.2114	&	1.5357	&	&	0.9454	&	0.8472	&	1.0492	&	1.2255	&	1.5759	\\
				
				AMML$_1$	&	0.9002	&	1.0152	&	1.0219	&	1.2870	&	1.4867	&	&	0.8160	&	0.8524	&	0.9379	&	1.3207	&	1.5742	\\
				
				RAMML$_1$	&	0.4557	&	0.4614	&	0.5554	&	0.9008	&	1.1409	&	&	0.0639	&	0.1018	&	0.0851	&	0.3059	&	0.5717	\\
				
				S	&	0.5249	&	0.6682	&	0.8161	&	1.4541	&	1.8625	&	&	0.5275	&	0.6597	&	0.7939	&	1.4660	&	1.8753	\\
				
				AMML$_2$	&	0.6607	&	0.9147	&	0.8968	&	1.2699	&	1.4605	&	&	0.3772	&	0.6000	&	0.6214	&	1.2966	&	1.5422	\\
				
				RAMML$_2$	&	0.2258	&	0.3525	&	0.3854	&	0.9031	&	1.1201	&	&	0.0642	&	0.1008	&	0.0847	&	0.2860	&	0.5390	\\
				\hline
				&	\multicolumn{11}{c}{$m=20$} \\
				\cline{2-12} 
				MM	&	0.5684	&	0.9978	&	0.9497	&	1.6374	&	1.9992	&	&	0.5490	&	1.0318	&	0.8953	&	1.6088	&	2.0397	\\
				
				LTS	&	1.3601	&	1.4243	&	1.5033	&	1.9556	&	2.5446	&	&	1.3503	&	1.4359	&	1.5317	&	1.8563	&	2.5387	\\
				
				AMML$_1$	&	1.0659	&	1.2139	&	1.2080	&	1.8028	&	2.2552	&	&	1.0480	&	1.1482	&	1.1866	&	1.7621	&	2.2977	\\
				
				RAMML$_1$	&	0.8478	&	0.9343	&	0.9641	&	1.5636	&	2.0235	&	&	0.1317	&	0.2179	&	0.1891	&	0.6689	&	1.2338	\\
				
				S	&	1.1751	&	1.8051	&	1.7165	&	2.4816	&	3.3106	&	&	1.1005	&	1.7525	&	1.6309	&	2.4502	&	3.3417	\\
				
				AMML$_2$	&	0.7525	&	1.1542	&	1.0870	&	1.7551	&	2.1846	&	&	0.6228	&	1.0910	&	0.9737	&	1.7253	&	2.2274	\\
				
				RAMML$_2$	&	0.5599	&	0.9062	&	0.8535	&	1.5359	&	1.9591	&	&	0.1334	&	0.2177	&	0.1893	&	0.6220	&	1.1623	\\
				\hline
				\hline
				\multicolumn{12}{c}{Contamination level = $20\%$} \\
				\hline
				&	\multicolumn{11}{c}{$m=1$} \\
				\cline{2-12} 
				&	\multicolumn{5}{c}{  $\ell=5$}   &	 & 	\multicolumn{5}{c}{  $\ell=10$} 			\\
				\cline{2-6}  \cline{8-12}
				MM	&	0.4461	&	0.9931	&	1.0734	&	3.0839	&	3.1299	&	&	0.3609	&	1.0414	&	0.9840	&	3.6322	&	3.7580	\\
				
				LTS	&	2.1464	&	1.7778	&	2.5918	&	3.1700	&	3.3282	&	&	2.3985	&	2.1247	&	2.7607	&	3.5390	&	3.8195	\\
				
				AMML$_1$	&	2.7559	&	2.8206	&	2.8515	&	2.9719	&	2.9715	&	&	2.8591	&	3.2785	&	3.3429	&	3.6997	&	3.6998	\\
				
				RAMML$_1$	&	0.1690	&	0.2314	&	0.2571	&	0.6988	&	0.8005	&	&	0.0073	&	0.0119	&	0.0100	&	0.0418	&	0.0688	\\
				
				S	&	0.4982	&	1.1499	&	1.2401	&	3.5014	&	3.5937	&	&	0.3781	&	1.1068	&	1.0245	&	3.7578	&	3.9076	\\
				
				AMML$_2$	&	2.4541	&	2.7179	&	2.6921	&	2.9850	&	2.9847	&	&	1.4806	&	2.6798	&	2.5184	&	3.7018	&	3.7049	\\
				
				RAMML$_2$	&	0.0543	&	0.1662	&	0.1431	&	0.6877	&	0.7970	&	&	0.0070	&	0.0119	&	0.0097	&	0.0398	&	0.0656	\\
				\hline
				&	\multicolumn{11}{c}{$m=5$} \\
				\cline{2-12} 
				MM	&	1.0223	&	1.0363	&	1.0332	&	1.0759	&	1.1504	&	&	1.0541	&	1.0680	&	1.0679	&	1.1116	&	1.1818	\\
				
				LTS	&	1.2814	&	1.2889	&	1.3084	&	1.2890	&	1.4834	&	&	1.3104	&	1.3122	&	1.3382	&	1.3256	&	1.4875	\\
				
				AMML$_1$	&	0.9921	&	1.0100	&	1.0050	&	1.0892	&	1.1774	&	&	1.0375	&	1.0525	&	1.0507	&	1.1288	&	1.2108	\\
				
				RAMML$_1$	&	0.7619	&	0.7933	&	0.7815	&	0.8871	&	0.9778	&	&	0.0405	&	0.0654	&	0.0573	&	0.1956	&	0.3214	\\
				
				S	&	1.3082	&	1.3170	&	1.3401	&	1.3314	&	1.5485	&	&	1.3187	&	1.3456	&	1.3467	&	1.3639	&	1.5658	\\
				
				AMML$_2$	&	0.9932	&	1.0109	&	1.0062	&	1.0835	&	1.1666	&	&	1.0382	&	1.0525	&	1.0513	&	1.1226	&	1.2012	\\
				
				RAMML$_2$	&	0.7670	&	0.7988	&	0.7873	&	0.8863	&	0.9722	&	&	0.0422	&	0.0669	&	0.0590	&	0.1870	&	0.3108	\\
				\hline
				&	\multicolumn{11}{c}{$m=10$} \\
				\cline{2-12} 
				MM	&	1.1484	&	1.1931	&	1.1811	&	1.2888	&	1.4655	&	&	1.1622	&	1.2124	&	1.1898	&	1.3184	&	1.4636	\\
				
				LTS	&	1.5638	&	1.5978	&	1.6118	&	1.7976	&	2.2175	&	&	1.5719	&	1.6271	&	1.6265	&	1.8212	&	2.2973	\\
				
				AMML$_1$	&	1.0943	&	1.1339	&	1.1165	&	1.3262	&	1.5427	&	&	1.1134	&	1.1600	&	1.1415	&	1.3402	&	1.5488	\\
				
				RAMML$_1$	&	0.9697	&	1.0133	&	0.9941	&	1.2034	&	1.4133	&	&	0.0925	&	0.1651	&	0.1362	&	0.4595	&	0.7448	\\
				
				S	&	1.6621	&	1.7289	&	1.7171	&	1.8785	&	2.4519	&	&	1.6672	&	1.7437	&	1.7504	&	1.9083	&	2.4676	\\
				
				AMML$_2$	&	1.0964	&	1.1357	&	1.1185	&	1.3084	&	1.5171	&	&	1.1153	&	1.1609	&	1.1433	&	1.3237	&	1.5173	\\
				
				RAMML$_2$	&	0.9760	&	1.0190	&	0.9998	&	1.1905	&	1.3934	&	&	0.0991	&	0.1732	&	0.1446	&	0.4428	&	0.7166	\\
				\hline
				&	\multicolumn{11}{c}{$m=20$} \\
				\cline{2-12} 
				MM	&	1.5137	&	1.5901	&	1.5740	&	1.7791	&	2.2969	&	&	1.4988	&	1.5764	&	1.5847	&	1.7845	&	2.2317	\\
				
				LTS	&	2.0562	&	2.2847	&	2.1948	&	3.0003	&	4.2073	&	&	2.0834	&	2.2013	&	2.2635	&	3.0191	&	4.2844	\\
				
				AMML$_1$	&	1.2698	&	1.3726	&	1.3359	&	1.8244	&	2.2852	&	&	1.2867	&	1.3659	&	1.3439	&	1.8106	&	2.2819	\\
				
				RAMML$_1$	&	1.2068	&	1.3077	&	1.2721	&	1.7406	&	2.1988	&	&	0.2928	&	0.4745	&	0.4030	&	1.1256	&	1.6735	\\
				
				S	&	2.2856	&	2.4812	&	2.4787	&	3.1068	&	4.5225	&	&	2.2711	&	2.4297	&	2.4752	&	3.1939	&	4.4752	\\
				
				AMML$_2$	&	1.2800	&	1.3773	&	1.3452	&	1.7765	&	2.2147	&	&	1.2944	&	1.3700	&	1.3521	&	1.7605	&	2.1994	\\
				
				RAMML$_2$	&	1.2214	&	1.3168	&	1.2862	&	1.7005	&	2.1384	&	&	0.3352	&	0.5164	&	0.4477	&	1.0955	&	1.6096	\\
				\hline
		\end{tabular}}
	\end{center}
\end{table}

\newpage

\begin{table}
	\caption{The simulated MSEs of MM, LMS, AMML$_1$, RAMML$_1$,   LTS, AMML$_2$, RAMML$_2$ estimators for scale parameter when the error distribution is $N(0,1)$.} \label{tab:simres3}
	\begin{center}
		\scriptsize{\begin{tabular}{l cccc c cccc}
				\hline
				&	\multicolumn{4}{c}{$\ell=5$}   &	 & 	\multicolumn{4}{c}{$\ell=10$} 			\\
				\cline{2-5} 	\cline{7-10}
				&	$m=1$	&	$m=5$	&	$m=10$	&	$m=25$	&	&	$m=1$	&	$m=5$	&	$m=10$	&	$m=25$	\\
				\hline\hline
				&\multicolumn{9}{c}{$n=50$}  \\
				\cline{2-10}
				&\multicolumn{9}{c}{Contamination level $=10\%$}  \\
				\cline{2-10}
				MM	&	0.0460	&	0.0233	&	0.0308	&	0.0860	&	&	0.0435	&	0.0238	&	0.0284	&	0.0866	\\
				
				LTS	&	0.0778	&	0.0619	&	0.0576	&	0.1296	&	&	0.0760	&	0.0702	&	0.0609	&	0.1243	\\
				
				AMML$_1$	&	0.3110	&	0.0777	&	0.0819	&	0.0689	&	&	0.0837	&	0.0706	&	0.0852	&	0.0655	\\
				
				RAMML$_1$	&	0.0382	&	0.0194	&	0.0301	&	0.0451	&	&	0.0406	&	0.0317	&	0.0325	&	0.0614	\\
				
				S	&	0.0460	&	0.0252	&	0.0329	&	0.0713	&	&	0.0435	&	0.0262	&	0.0298	&	0.0734	\\
				
				AMML$_2$	&	0.2623	&	0.0660	&	0.0626	&	0.0673	&	&	0.0621	&	0.0497	&	0.0505	&	0.0645	\\
				
				RAMML$_2$	&	0.0390	&	0.0251	&	0.0283	&	0.0378	&	&	0.0410	&	0.0334	&	0.0353	&	0.0502	\\
				\hline
				&	\multicolumn{9}{c}{Contamination level $=20\%$}  \\
				\cline{2-10}
				MM	&	0.1366	&	0.0618	&	0.1897	&	0.3920	&	&	0.1466	&	0.0577	&	0.1875	&	0.4297	\\
				
				LTS	&	0.3193	&	0.0786	&	0.2786	&	0.7529	&	&	0.3655	&	0.0689	&	0.2716	&	0.7505	\\
				
				AMML$_1$	&	0.6533	&	0.0418	&	0.0421	&	0.6658	&	&	0.5751	&	0.0484	&	0.0472	&	0.6669	\\
				
				RAMML$_1$	&	0.0411	&	0.0183	&	0.0503	&	0.6979	&	&	0.0537	&	0.0491	&	0.0865	&	0.6993	\\
				
				S	&	0.1366	&	0.0570	&	0.1498	&	0.2782	&	&	0.1467	&	0.0544	&	0.1479	&	0.3162	\\
				
				AMML$_2$	&	0.6240	&	0.0433	&	0.0303	&	0.0890	&	&	0.4141	&	0.0478	&	0.0321	&	0.1075	\\
				
				RAMML$_2$	&	0.0367	&	0.0188	&	0.0231	&	0.1093	&	&	0.0555	&	0.0481	&	0.0556	&	0.1378	\\
				\hline
				\hline
				&\multicolumn{9}{c}{$n=200$}  \\
				\cline{2-10}
				&\multicolumn{9}{c}{Contamination level $=10\%$}  \\
				\cline{2-10}
				MM	&	0.0282	&	0.0211	&	0.0181	&	0.0129	&	&	0.0181	&	0.0225	&	0.0187	&	0.0132	\\
				
				LTS	&	0.0571	&	0.0751	&	0.0717	&	0.0565	&	&	0.0717	&	0.0801	&	0.0750	&	0.0599	\\
				
				AMML$_1$	&	0.3020	&	0.0771	&	0.0704	&	0.0689	&	&	0.0704	&	0.0613	&	0.0593	&	0.0721	\\
				
				RAMML$_1$	&	0.0294	&	0.0093	&	0.0175	&	0.0256	&	&	0.0175	&	0.0235	&	0.0224	&	0.0204	\\
				
				S	&	0.0282	&	0.0219	&	0.0186	&	0.0113	&	&	0.0186	&	0.0229	&	0.0196	&	0.0124	\\
				
				AMML$_2$	&	0.2522	&	0.0686	&	0.0500	&	0.0389	&	&	0.0500	&	0.0291	&	0.0201	&	0.0342	\\
				
				RAMML$_2$	&	0.0303	&	0.0160	&	0.0196	&	0.0227	&	&	0.0196	&	0.0251	&	0.0249	&	0.0240	\\
				\hline
				&	\multicolumn{9}{c}{Contamination level $=20\%$}  \\
				\cline{2-10}
				MM	&	0.1393	&	0.0115	&	0.0236	&	0.0768	&	&	0.1364	&	0.0119	&	0.0239	&	0.0728	\\
				
				LTS	&	0.4019	&	0.0182	&	0.0110	&	0.0295	&	&	0.4286	&	0.0212	&	0.0115	&	0.0260	\\
				
				AMML$_1$	&	0.6431	&	0.0375	&	0.0381	&	0.0310	&	&	0.6230	&	0.0413	&	0.0400	&	0.0320	\\
				
				RAMML$_1$	&	0.0133	&	0.0071	&	0.0106	&	0.0093	&	&	0.0495	&	0.0352	&	0.0292	&	0.0157	\\
				
				S	&	0.1393	&	0.0115	&	0.0230	&	0.0753	&	&	0.1364	&	0.0118	&	0.0237	&	0.0730	\\
				
				AMML$_2$	&	0.6317	&	0.0339	&	0.0331	&	0.0238	&	&	0.3894	&	0.0375	&	0.0347	&	0.0247	\\
				
				RAMML$_2$	&	0.0164	&	0.0061	&	0.0086	&	0.0067	&	&	0.0490	&	0.0368	&	0.0308	&	0.0168	\\
				\hline
		\end{tabular}}
	\end{center}
\end{table}

\newpage

\begin{table}[t]
	\caption{The simulated MSEs of MM, LMS, AMML$_1$, RAMML$_1$,   LTS, AMML$_2$, RAMML$_2$ estimators for regression coefficients when the distribution of error terms is $N(0,1)$ with zero contamination.} \label{tab:simres4}
	\begin{center}
		\footnotesize{\begin{tabular}{l cccc c cccc}
				\hline
				&	\multicolumn{4}{c}{ $n=50$}   &	 & 	\multicolumn{4}{c}{ $n=200$} 			\\
				\cline{2-5}  \cline{7-10}
				Estimator&$m=1$	&	$m=5$	&	$m=10$	&	$m=20$	&& $m=1$	&	$m=5$	&	$m=10$	&	$m=20$	\\
				\hline
				LS	&	0.0222	&	0.1145	&	0.2600	&	0.7357	&	&	0.0051	&	0.0251	&	0.0531	&	0.1112	\\
				
				MM	&	0.0231	&	0.1249	&	0.2922	&	1.0148	&	&	0.0053	&	0.0270	&	0.0590	&	0.1287	\\
				
				LTS	&	0.0344	&	0.2306	&	0.5866	&	1.8022	&	&	0.0074	&	0.0376	&	0.0803	&	0.1793	\\
				
				AMML$_1$	&	0.0224	&	0.1166	&	0.2636	&	0.7574	&	&	0.0051	&	0.0254	&	0.0542	&	0.1135	\\
				
				RAMML$_1$	&	0.0273	&	0.1180	&	0.2654	&	0.7635	&	&	0.0065	&	0.0255	&	0.0546	&	0.1143	\\
				
				S	&	0.0663	&	0.4081	&	0.9027	&	2.2393	&	&	0.0162	&	0.0955	&	0.2187	&	0.4537	\\
				
				AMML$_2$	&	0.0224	&	0.1172	&	0.2653	&	0.7628	&	&	0.0051	&	0.0255	&	0.0544	&	0.1141	\\
				
				RAMML$_2$	&	0.0274	&	0.1187	&	0.2677	&	0.7702	&	&	0.0065	&	0.0256	&	0.0548	&	0.1149	\\
				\hline
		\end{tabular}}
	\end{center}
\end{table}

\begin{table}[t]
	\caption{The simulated MSEs of MM, LMS, AMML$_1$, RAMML$_1$,   LTS, AMML$_2$, RAMML$_2$ estimators for regression coefficients with correlated predictors: $n=200$, $\ell = 10$ and contamination level $=10\%$.} \label{tab:simres.correlated}
	\begin{center}
		\scriptsize{\begin{tabular}{l ccccc }
				\hline
				Estimator	&	Normal	&	Laplace	&$t_5$	&	$t_1$&	Slash	\\
				\hline
				&	\multicolumn{5}{c}{$m=5$} \\
				\cline{2-6} 
				MM	&	0.2415	&	0.3600	&	0.3311	&	0.8236	&	2.0186	\\
				
				LTS	&	0.2822	&	0.3797	&	0.3697	&	0.7139	&	2.6958	\\
				
				AMML$_1$	&	0.2590	&	0.5424	&	0.4623	&	2.9306	&	4.1066	\\
				
				RAMML$_1$	&	0.2447	&	0.3780	&	0.3460	&	1.2012	&	2.1839	\\
				
				S	&	0.6538	&	0.4250	&	0.6039	&	0.5792	&	2.9211	\\
				
				AMML$_2$	&	0.2540	&	0.4548	&	0.4121	&	2.5534	&	3.9090	\\
				
				RAMML$_2$	&	0.2455	&	0.3723	&	0.3440	&	1.1143	&	2.0515	\\
				\hline
				&	\multicolumn{5}{c}{$m=10$} \\
				\cline{2-6} 
				MM	&	0.5865	&	1.1378	&	0.8459	&	8.5211	&	14.4712	\\
				
				LTS	&	1.8531	&	5.5060	&	5.7911	&	27.7480	&	44.4990	\\
				
				AMML$_1$	&	0.9999	&	2.8892	&	2.4072	&	10.8070	&	13.7108	\\
				
				RAMML$_1$	&	0.6063	&	1.0319	&	0.8598	&	3.4661	&	5.9173	\\
				
				S	&	2.2357	&	2.7599	&	2.5881	&	34.5933	&	49.7293	\\
				
				AMML$_2$	&	0.8313	&	1.8118	&	1.4004	&	10.5471	&	13.3625	\\
				
				RAMML$_2$	&	0.6079	&	1.0161	&	0.8540	&	3.2843	&	5.5056	\\
				\hline
				&	\multicolumn{5}{c}{$m=20$} \\
				\cline{2-6} 
				MM	&	1.2732	&	1.9421	&	1.7584	&	4.8325	&	9.6923	\\
				
				LTS	&	1.4942	&	2.0848	&	2.0331	&	4.3235	&	9.7048	\\
				
				AMML$_1$	&	1.1957	&	2.0302	&	1.7448	&	8.5954	&	22.3637	\\
				
				RAMML$_1$	&	1.2593	&	2.0304	&	1.7781	&	6.6322	&	12.7007	\\
				
				S	&	4.3629	&	3.4980	&	6.5179	&	21.3058	&	48.5781	\\
				
				AMML$_2$	&	1.3343	&	2.1102	&	2.1823	&	11.3807	&	25.6140	\\
				
				RAMML$_2$	&	1.2640	&	1.9994	&	1.7705	&	6.6776	&	12.9944	\\
				\hline
		\end{tabular}}
	\end{center}
\end{table}

\begin{table}[h]
	\caption{The estimated values of the regression coefficients and the scale parameter and also SEP and SEP$_{\text{trim}}$  values for the starsCYG data.} \label{tab:starCYGestimates}
	\begin{center}
		\small{	\begin{tabular}{ll cccc cc}
				\hline
				Method && $\beta_0$ &$\beta_1$  & $\sigma$ && SEP & 	SEP$_{\text{trim}}$ \\
				\cline{1-1} 	\cline{3-5}  	\cline{7-8} 
				MM	&&	-4.9694	&	2.2532	&	0.4715	&	&	0.9556	&	0.3282	\\
				
				LTS	&&	-8.5001	&	3.0462	&	0.4562	&	&	1.1507	&	0.3266	\\
				
				AMML$_1$	&&	3.4697	&	0.3409	&	0.5239	&	&	0.6000	&	0.3994	\\
				
				RAMML$_1$	&&	-8.0822	&	2.9523	&	0.3249	&	&	1.1269	&	0.3252	\\
				
				S	&&	-9.5708	&	3.2904	&	0.4715	&	&	1.2133	&	0.3325	\\
				
				AMML$_2$	&&	3.3016	&	0.3790	&	0.5214	&	&	0.6041	&	0.3972	\\
				
				RAMML$_2$	&&	-8.0907	&	2.9553	&	0.3249	&	&	1.1277	&	0.3252	\\
				\hline
		\end{tabular}}
	\end{center}
\end{table}

\begin{table}[h]
	\caption{The estimated regression coefficients, scale parameter and values of the performance criteria for 
		the Aircraft data.} \label{tab:aircraft}
	\begin{center}
		\footnotesize{\begin{tabular}{l ccccccccc cc}
				\hline
				Method	&	&	$\beta_0$	&	$\beta_1$	&	$\beta_2$	&	$\beta_3$	&	$\beta_4$	& $\sigma$&	&	SEP	&	SEP$_{\text{trim}}$		\\
				\cline{1-1} 	\cline{3-8}  	\cline{10-11} 
				MM	&	&	6.1396	&	-3.2306	&	1.6713	&	0.0019	&	-0.0009	&	5.8932	&	&	10.5866	&	3.4636	\\
				
				LTS	&	&	9.5007	&	-3.0488	&	1.2100	&	0.0014	&	-0.0006	&	5.6927	&	&	12.4080	&	3.4110	\\
				
				AMML$_1$	&	&	2.2997	&	-3.4300	&	2.0052	&	0.0025	&	-0.0013	&	6.5780	&	&	8.9124	&	3.8884	\\
				
				RAMML$_1$	&	&	7.0264	&	-3.2411	&	1.6622	&	0.0018	&	-0.0009	&	3.9639	&	&	11.1015	&	3.4651	\\
				
				S	&	&	13.3733	&	-4.0220	&	1.5413	&	0.0017	&	-0.0010	&	5.8932	&	&	11.8666	&	3.5001	\\
				
				AMML$_2$	&	&	1.8030	&	-3.4535	&	2.0451	&	0.0026	&	-0.0014	&	6.7367	&	&	8.7372	&	3.9778	\\
				
				RAMML$_2$	&	&	6.9195	&	-3.2381	&	1.6674	&	0.0018	&	-0.0009	&	3.9821	&	&	11.0630	&	3.4699	\\
				\hline
		\end{tabular}}
	\end{center}
\end{table}

\clearpage

\begin{thebibliography}{99}
	

\bibitem{Donmez2010}  Donmez A. Adaptive estimation and hypothesis testing methods. METU; 2010.
	
\bibitem{RousseeuwLeroy1987}  Rousseeuw PJ, Leroy AM. Robust Regression and Outlier Detection. Wiley; 1987.	

\bibitem{Huber1973}  Huber PJ. Robust regression: Asymptotics, conjectures, and monte carlo. Ann. Stat. 1973;1:799-821.

\bibitem{RousseeuwYohai1984}  Rousseeuw P, Yohai V. Robust Regression by Means of S-Estimators. Robust nonlinear time Ser. Anal. Springer; 1984. p. 256-272.

\bibitem{Rousseeuw1984}  Rousseeuw PJ. Least median of squares regression. J. Am. Stat. Assoc. 1984;79:871-880.

\bibitem{Yohai1987} Yohai VJ. High Breakdown-Point and High Efficiency Robust Estimates for Regression. Ann. Stat. 1987;15:642-656.

\bibitem{Maronnaetal2006}  Maronna RA, Martin RD, Yohai VJ. Robust Statistics: Theory and Methods. Robust Stat. Theory Methods. Wiley; 2006.

\bibitem{GschwandtnerFilzmoser2012}  Gschwandtner M, Filzmoser P. Computing Robust Regression Estimators: Developments since Dutter 1977. Austrian J. Stat. 2012;41:45-58.

\bibitem{YuYao2017}  Yu C, Yao W. Robust linear regression: A review and comparison. Commun. Stat. Simul. Comput. Taylor and Francis Inc.; 2017. p. 6261-6282.

\bibitem{Tiku1967}  Tiku ML. Estimating the mean and standard deviation from a censored normal sample. Biometrika. 1967;54:155-1665.

\bibitem{Tiku1968}  Tiku ML. Estimating the parameters of normal and logistic distributions from a censored normal sample. Aust. J. Stat. 1968;10:64-74.

\bibitem{TikuSuresh1992}  Tiku ML, Suresh RP. A new method of estimation for location and scale parameters. J. Stat. Plan. Inference. 1992;30:281-292.

\bibitem{Tikuetal2001}  Tiku ML, Islam MQ, Selcuk AS. Nonnormal regression. II. Symmetric distributions. Commun. Stat. - Theory Methods. 2001;30:1021-1045.

\bibitem{IslamTiku2005}  Islam MQ, Tiku ML. Multiple Linear Regression Model Under Nonnormality. Commun. Stat. - Theory Methods. 2005;33:2443-2467.


\bibitem{Lucas1997}  Lucas A. Robustness of the student t based M-estimator. Commun. Stat. - Theory Methods. 1997;26:1165-1182.

\bibitem{ArslanGenc2003}  Arslan O, Genc AI. Robust location and scale estimation based on the univariate generalized t (GT) distribution. Commun. Stat. Methods. 2003;32:1505-1525.


\bibitem{ArslanGenc2009}  Arslan O, Genc AI. The skew generalized t distribution as the scale mixture of a skew exponential power distribution and its applications in robust estimation. Statistics (Ber). 2009;43:481-498.

\bibitem{Acitasetal2013a}  Acitas S, Kasap P, Senoglu B, et al. Robust estimation with the skew t2 distribution. Pakistan J. Stat. 2013;29:409-430.


\bibitem{Acitasetal2013b}  Acitas S, Kasap P, Senoglu B, et al. One-step M-estimators: Jones and Faddy’s skewed t-distribution. J. Appl. Stat. 2013;40:1545-1560.

\bibitem{TikuSurucu2012}  [1] Tiku ML, Sürücü B. MMLEs are as good as M-estimators or better. Stat. Probab. Lett. 2009;79:984-989.

\bibitem{Acitasetal2019}  Acitas S, Filzmoser P, Senoglu B. A Robust Adaptive Modified Maximum Likelihood Estimator for the Linear Regression Model. Olomoucian Days Appl. Math. 2019. Olomouc, Czech Republic; 2019.

\bibitem{PuthenpuraSinha1986}  Puthenpura S, Sinha NK. Modified maximum likelihood method for the robust estimation of system parameters from very noisy data. Automatica1. 1986;22:231-235.

\bibitem{Vaughan1992} Vaughan DC. On the Tiku--Suresh method of estimation. Commun. Stat. - Theory Methods. 1992;21:329-340.


\bibitem{Bhattacharyya1985}  Bhattacharyya GK. The asymptotics of maximum likelihood and related estimators based on type ii censored data. J. Am. Stat. Assoc. 1985;80:398-404.

\bibitem{Lopuhaaetal1991}  Lopuhaa HP, Rousseeuw PJ. Breakdown points of affine equivariant estimators of multivariate location and covariance matrices. Ann. Stat. 1991;19:229-248.

\bibitem{Crouxetal2007}  Croux C, Filzmoser P, Oliveira MR. Algorithms for Projection-Pursuit robust principal component analysis. Chemom. Intell. Lab. Syst. 2007;87:218-225.

\bibitem{Fritzetal2012} Fritz, H., Filzmoser P, Croux. C. A comparison of algorithms for the multivariate $L_1$-median. Comput. Stat. 2012;27:393-410.

\bibitem{Liebmannetal2010} Liebmann B, Filzmoser P, Varmuza K. Robust and classical PLS regression compared. J. Chemom. 2010;24:111-120.

\bibitem{Filzmoseretal2011}  Filzmoser P, Todorov V. Review of robust multivariate statistical methods in high dimension. Anal. Chim. Acta. 2011;705:2-14.


	
\bibitem{Gray1985}  Gray JB. Graphics for Regression Diagnostics. Am. Stat. Assoc. Proc. Stat. Comput. Sect. ASA. Washington, D.C.,U.S.A.; 1985.



















	











\end{thebibliography}
\end{document}